\documentclass[aps,pra,twocolumn,showpacs,superscriptaddress,amssymb,floatfix,tightenlines]{revtex4-1}

\usepackage{epsfig}
\usepackage{dcolumn}
\usepackage{bm}
\input{epsf}
\usepackage{bbold}
\usepackage{verbatim}
\usepackage{dsfont} 
\usepackage{amsmath,amssymb,bm}
\usepackage{graphicx}
\usepackage{amsthm}
\usepackage{hyperref} 
\usepackage{color}

\usepackage[nottoc]{tocbibind}

\newcommand{\ket}[1]{\left| #1 \right\rangle}
\newcommand{\bra}[1]{\left\langle #1 \right|}
\newcommand{\proj}[1]{\ket{#1}\bra{#1}}

\newcommand{\ketbra}[2]{| #1 \rangle\langle #2 |}

\begin{document}
\title{Conditional quantum nonlocality in dimeric and trimeric arrays of organic molecules}
\author{John H. Reina}
\altaffiliation{{\tt john.reina@correounivalle.edu.co}}
\address{Centre for Bioinformatics and Photonics---CIBioFi, Calle 13 No.~100-00, Edificio 320 No.~1069, Universidad del Valle, 760032 Cali, Colombia}
\address{Departamento de F\'isica, Universidad del Valle, 760032 Cali, Colombia}
\author{Cristian E. Susa}
\altaffiliation{\tt cristiansusa@correo.unicordoba.edu.co}
\address{Centre for Bioinformatics and Photonics---CIBioFi, Calle 13 No.~100-00, Edificio 320 No.~1069, Universidad del Valle,  760032 Cali, Colombia}
\address{Departamento de F\'isica y Electr\'onica, Universidad de C\'ordoba, 
230002 Monter\'ia, Colombia}
\author{Richard Hildner}
\address{Soft Matter Spectroscopy, Universit\"at Bayreuth, Universit\"atsstrasse 30, 95447 Bayreuth, Germany}

\begin{abstract}
{Arrays of covalently bound organic molecules possess potential for light-harvesting and energy transfer applications due to the strong coherent dipole-dipole coupling between the transition dipole moments of the molecules involved. Here, we show that such molecular systems, based on perylene-molecules, can be considered as arrays of qubits that are amenable for laser-driven quantum coherent control. The perylene monomers exhibit dephasing times longer than  four orders of magnitude a typical gating time, thus allowing for the execution of a large number of  gate operations on the sub-picosecond timescale. Specifically, we demonstrate quantum logic gates and entanglement in bipartite (dimer) and tripartite (trimer) systems of perylene-based arrays. In dimers, naturally entangled states with a tailored degree of entanglement can be produced. The nonlocality of the molecular trimer entanglement is demonstrated by testing Mermin's (Bell-like) inequality violation.
}
\end{abstract}


\maketitle

\section{Introduction}
\label{intro}
Quantum coherence has been identified as an emergent resource~\cite{richard2015,scholes-nat17,jh-sr17,scholes-mat17,reina2009,pleniormp,reina2009-2} 
for biological and chemical functionality~\cite{scholes-nat17,pleniormp}. Understanding and, particularly, exploiting these features on a molecular level has become feasible in recent years through the progress in spectroscopy and quantum control of single molecular systems~\cite{Basche,Weigel2015,Brinks2014,Accanto2016,Hildner:2011dn}. Recent evidence points out that quantum coherence can be robust and survive even at ambient conditions~\cite{richard2015,scholes-mat17,scholes-nat17,romero17,reina2009,pleniormp,reina2009-2,scholes-rev11,huelga13}, a fact that can be 
harnessed for engineering and transferring quantum information in a wide variety of organic nanosystems: Multichromophoric and biomolecular structures for light harvesting~\cite{scholes-rev11,huelga13,reina2009,pleniormp,reina2009-2,kohler06,Issac,romero17}, as well as complex chemical structures for organic photovoltaics with relevance to sustainable renewable energy production~\cite{scholes-mat17,romero17,scholes-nat17,polman16}. An important advantage of organic systems is that these materials can be easily scaled up by chemical synthesis~\cite{richard2015,polman16,scholes-mat17}, and do not require complex settings like high-vacuum traps for their implementation~\cite{polman16,scholes-mat17,romero17}.  

Here, we show that, thanks to  the recent advances in single-molecule spectroscopy, we are able to
manipulate and to individually control molecular 
dynamics 
on the picosecond and sub-picosecond time scales, i.e., we can 
generate a conditional  coherent  quantum dynamics and robust entanglement 
in Perylene-Bisimide (PBI) based arrays immersed in an organic matrix. 
We specifically focus on such polycyclic aromatic hydrocarbon based molecules, because they can be easily synthesised and can be externally driven with a high degree of control~\cite{Hildner2007,Issac,Issac2012,Weigel2015,Hildner:2011dn}.

This paper is organised as follows: In section~\ref{pbiarray}, we briefly introduce the physical properties of 
the PBI dimer and trimer according to spectroscopy data. Section~\ref{pbitheory}
describes the theory behind the temporal evolution of  dimer and trimer states.  Dimer's structure for implementing quantum logic gates and entanglement is shown in Section~\ref{dimercoherence}. We describe entanglement generation and quantum nonlocality in trimers in Section~\ref{trimercase}. Finally, a summary of our findings and experimental remarks for a physical  implementation are discussed in Section~\ref{summary}.
\begin{figure}[h]
  \centering
  \includegraphics[width=\columnwidth]{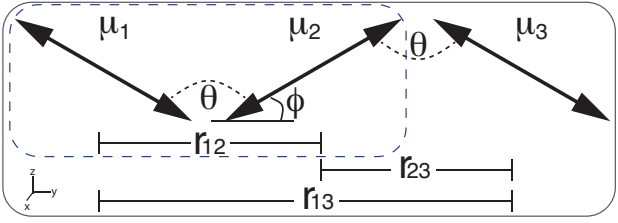}
  \caption{Schematics of the mutual orientations of the transition dipole moments (double-headed arrows) of covalently bound PBI molecules in trimer (black-solid box) and dimer arrangement (blue-dashed box). $\bm{\mu}_i$ corresponds to the transition dipole moment of the $i$-th PBI molecule.
  $\theta$ is the angle between subunits 1 and 2 (also 2 and 3). Subunits 1 and 3 are parallel to each other. The separation vectors between PBI molecules are $|\bm{r}_{12}|=|\bm{r}_{23}|=|\bm{r}_{13}|/2$.}
  \label{DDS}
\end{figure}

\section{Spectroscopy of molecular dimers and trimers: defining molecular qubit registers}
\label{pbiarray}

As building blocks for quantum information processing units we consider here a molecular array consisting of two (three)  PBI molecules that are covalently linked by a rigid calix[4]arene bridge~\cite{Ernst,Issac}. In the following those arrays will be referred to as dimers (trimers). We specifically focus on those PBI systems because we characterised their photophysics extensively by single-molecule techniques~\cite{Issac,Issac2012,Hildner2007,Ernst}; moreover, PBIs are very bright and photostable. Considering for each PBI molecule only the lowest-energy optical transition, i.e., the transition between the electronic ground state ($\ket{g}$) and the vibrationless lowest-energy excited state ($\ket{e}$), each PBI in a dimer (trimer) represents a two-state (qubit) system, with basis states $\ket{g_i}\equiv \ket{0_i}$ and $\ket{e_i}\equiv \ket{1_i}$. Thus, in what follows, $\{\ket{0_i},\ket{1_i}\}$ denotes the 
computational basis associated to qubit $i$-th. 

The transition dipole moment $\bm{\mu}_i$ ($i=1,2,3$) for this lowest-energy transition is oriented along the long axis of PBI. Owing to the rigid bridge the zig-zag-type arrangement for the transition dipole moments shown in Fig.~\ref{DDS} results for a dimer (trimer), with a centre-to-centre distance of $|\bm{r}_{12}|=|\bm{r}_{23}|=2.2$~nm and an opening angle $\theta=2\pi/3$. 

For the specific quantum control experiments proposed here 
we consider dimers and trimers embedded in a well-defined, crystalline matrix at cryogenic temperature (1.5 K). Under these conditions we found that the lowest-energy optical transition in PBI molecules occurs at a photon frequency of 
$\nu_{i} \sim 522$~THz, corresponding to a wavelength of $\lambda_{i}=575$~nm 
~\cite{Hildner2007,Issac2012}. 
Moreover, in this situation the homogeneous line width of the PBI molecules~\cite{Basche},
$
	\gamma_h=1/(2\pi T_1)+1/(\pi T_2^{*})
$,
is entirely determined by its excited state lifetime $T_1$, because pure dephasing processes, described by the time constant $T_2^{*}$, are frozen out ($T_2^{*} \rightarrow \infty$). For PBI molecules we measured $T_1=5.8$ ns~\cite{Hildner2007}, and thus we obtain $\gamma_h \sim 27$ MHz. 

A further fundamental physical parameter for our dimer and trimer systems is the 
nearest-neighbour electronic coupling $V_{ij} (i \neq j)$ between the transition dipole moments of the individual PBI molecules. Given the magnitude of the transition dipole moment $|\bm{\mu}_i| = 10$ D~\cite{Hildner2007,Ernst} and the relatively small centre-to-centre distance, an electronic coupling of $\sim1356$~GHz (or $45$~cm$^{-1}$) between adjacent PBIs can be calculated (see Appendix A). 

An important figure of merit for performing quantum gates on a dimer (trimer) is the ratio between the nearest-neighbour electronic coupling $V_{ij}$ and the molecular detuning, which is defined as $\Delta_{ij} := \nu_i - \nu_j$. 
Previously, we considered a ratio of $V_{ij}/\Delta_{ij}\sim0.1$ as typical 
for performing dimeric conditional quantum dynamics~\cite{jh04}. However, we found experimentally that the difference in transition frequencies $\Delta_{ij}$ can assume any value between $0$ and $570$ cm$^{-1}$ ($17$~THz) depending on the specific local environment for each PBI molecule in a dimer (trimer)~\cite{Issac,Issac2012}, even if embedded in a well-defined matrix at low temperatures. This means that the ratio $V_{ij}/\Delta_{ij}$ can run from very large ($\gg 1$) to small ($\ll 1$), depending on the specific dimer (trimer) under investigation. As we cannot control this detuning experimentally,
we perform initial calculations for some exemplary values in the entire range ($\gg 1$ to $\ll 1$).
Then we will proceed  to identify which effects are to be expected and what to look for in the
experiments. Since both $V_{ij}$ and $\Delta_{ij}$ are much smaller than the transition frequencies of the PBI molecules, the rotating wave approximation (RWA) is well suited for describing  the dimers' and trimers' quantum dynamics~\cite{jh04,susa2012}.

\section{Dimer and trimer dissipative quantum dynamics}
\label{pbitheory}

For the mathematical description of the quantum dynamics of PBI dimers and trimers we follow the description given 
in~\cite{jh04,Hettich,susa2010,susa2012,jh14,Ficekbook}. For the dimer, the effective Hamiltonian after making the standard Born-Markov approximation on the 
system-environment interaction~\cite{susa2012,Ficekbook,jh14}, can be written as ($h=1$),
\begin{equation}
	H_{\mathrm{dimer}} = H_{Q}+H_{12} ,
\end{equation} 
where 
$H_Q=-\frac{1}{2}(\nu_1\sigma^{(1)}_{z}+\nu_2\sigma^{(2)}_{z})$, 
and ${H}_{12}~=~\frac{1}{2} V_{12}( 
	\sigma^{(1)}_{x}\otimes\sigma^{(2)}_{x} 
	+ \sigma^{(1)}_{y}\otimes\sigma^{(2)}_{y})$.

The matrix representation of $H_{\mathrm{dimer}}$ in the computational basis of product states $\ket{i_1}\otimes\ket{j_2}$ ($i,j = 0,1$) reads
\begin{eqnarray}
	\label{hrwa}
	H_{\mathrm{dimer}}= 
	\left(
	\begin{array}{cccc}
		-\nu_0  & 0 & 0 & 0 \\
		0 & -\frac{\Delta_-}{2}  & V_{12} & 0 \\
		0 & V_{12} & \frac{\Delta_-}{2} & 0 \\
		0 & 0 & 0 & \nu_0
	\end{array}\right) ,
\end{eqnarray}
where the molecular detuning is
$\Delta_-:=\Delta_{12}=\nu_1-\nu_2$, and $2\nu_0=\nu_1+\nu_2$.

An external control can be included to the dynamics by means of the light-matter Hamiltonian 
${H}_{L}=\Omega_{i}/2(\sigma^{(i)}_{-}e^{\mathrm{i}\omega_L t}+\sigma^{(i)}_{+}e^{-\mathrm{i}\omega_L t})$~\cite{jh04,susa2012,jh14}, 
$\omega_L=2\pi\nu_L$, where $\nu_L$ denotes the laser frequency, and
$\Omega_{i}=-\boldsymbol{\mu}_i\cdot \boldsymbol{E}_i$ gives the 
Rabi frequency induced by the interaction between the $i$-th transition dipole moment $\bm{\mu}_{i}$ and the coherently driving electric field 
$\boldsymbol{E}_i$ acting on qubit $i$ located at position $\bm{r}_i$. 
$\sigma_{+}^{(i)}=\ket{1_{i}}\bra{0_{i}}$ and 
$\sigma_{-}^{(i)}=\ket{0_{i}}\bra{1_{i}}$ stand for the raising and lowering operators, respectively.  Due to the short separation between qubits compared to the optical diffraction limit, we consider that the 
laser affects both qubits in the same way. Hence, in our simulations we fix $\Omega_{1}=\Omega_{2}=\Omega$.

Since we consider here cryogenic temperatures, we can assume a zero-temperature environment. Within the weak light-matter interaction (Born-Markov) approximation, the time evolution of the 
density matrix operator associated to the qubit-qubit system 
can then be approached by means of the quantum master equation~\cite{Ficekbook,jh04}
\begin{eqnarray}
  \label{ME}
  {\dot\rho}&=& - \text{i}  [ \tilde{H}_{\mathrm{dimer}}, {\rho} ]\\\nonumber
  &&- 
  \frac{1}{2}\sum_{i,j=1}^{2}\Gamma _{ij} \Big( 
    {\rho} \sigma_{+}^{(i)} \sigma_{-}^{(j)} + 
    \sigma_{+}^{(i)}\sigma_{-}^{(j)}{\rho}
    -2\sigma_{-}^{(j)}{\rho} \sigma_{+}^{(i)} 
  \Big),
\end{eqnarray}
where $\tilde{H}_{\mathrm{dimer}}=H_{\mathrm{dimer}}+H_L$. The density matrix elements are denoted by $\rho_{ij,kl}$, with $i,j,k,l=0,1$. 
$\Gamma_{ii}\equiv\Gamma$ are the spontaneous emission rates, and $\Gamma_{ij}, i\neq j,$ represent cross-damping rates, for which the explicit forms are given in appendix \ref{colleffects}.
Given the PBI excited state lifetime of $T_1 \sim 5.8$ ns, we get $\Gamma_{i}=1/T_1 \sim172$~MHz. 
Based on Eq.~\eqref{CDR} we estimate the cross-damping rate to $\Gamma_{12}\sim -86$~MHz.

For the trimer we are able to derive analytical expressions for the three-qubit eigensystem by considering
that qubit 1 and qubit 3 (the `outer' PBI molecules, see Fig.~\ref{DDS}) have the same transition frequency $\nu$. 
Hence, the only molecular detuning reads $\Delta_-:=\nu_2-\nu$ ($\Delta_{21}=\Delta_{23}$), where $\nu_2\,$ is the transition frequency of the `middle' qubit. Here, for the ease of notation, the same symbol $\Delta_-$ as for the dimer case is used, but we should be aware that its 
definition is different. 
Due to the spatial symmetry of the trimer shown in Fig.~\ref{DDS}, we also 
have $V_{12}=V_{23}\equiv V$, and $V>V_{13}$. Under this consideration, and without loss of generality, the effective 
three-qubit bare Hamiltonian can now be written as
\begin{eqnarray}
	\label{trimersystem}
	&& H_{\mathrm{trimer}}= \\ \nonumber
	&& 
	\begin{small}
	\left(
	\begin{array}{cccccccc}
		-\frac{3\nu_0}{2}  & 0 & 0 & 0 &  & 0 & 0 & 0 \\
		0 & -\frac{\nu_2}{2}  & V & 0 & V_{13} & 0 & 0 & 0 \\
		0 & V & -\frac{\nu-\Delta_-}{2} & 0 & V & 0 & 0 & 0\\
		0 & 0 & 0 & \frac{\nu_2}{2} & 0 & V & V_{13} & 0 \\
		0 & V_{13} & V & 0 & -\frac{\nu_2}{2}  & 0 & 0 & 0 \\
		0 & 0 & 0 & V & 0 & \frac{\nu-\Delta_-}{2} & V & 0 \\
		0 & 0 & 0 & V_{13} & 0 & V & \frac{\nu_2}{2} & 0 \\
		0 & 0 & 0 & 0 & 0 & 0 & 0 & \frac{3\nu_0}{2}
	\end{array}\right)
\end{small}
\end{eqnarray}
where $\nu_0=(\nu_1+\nu_2+\nu_3)/3=(2\nu+\nu_2)/3$, 
and $\Delta_-=\nu_2-\nu\ll\nu$. The dynamics 
of the trimer system is described by a master equation similar to that of Eq.~\eqref{ME}, but replacing  
$\tilde{H}_{\mathrm{dimer}}$ by $H_{\mathrm{trimer}}$  (the total trimer Hamiltonian with the laser action 
is given in Appendix~\ref{TQHam}), and by extending  the incoherent sum term over  indexes $i,j$ from $1$ to $3$.

\section{PBI dimer quantum coherence and logic gating}
\label{dimercoherence}

For  the dimer we next illustrate how one- and two-qubit logic gates, and hence entanglement and nonlocal correlations generation, is achieved. 
The dynamics of the dimer (two-qubit) system is described by means of the master equation~\eqref{ME}, 
from which we obtain the density matrix and are able to simulate the physical realisation of logic gates as well as the generation of entanglement. 

According to our description in the previous sections, the spontaneous emission rate ($\sim$ 200 MHz) of PBI is up to five orders of magnitude smaller than the electronic coupling $V_{12}$ and the molecular detuning $\Delta_-$ ($10^3$  and $10^4$~GHz, respectively). Since emission is the only dissipation channel, the dimer is a highly coherent quantum system. As we will show below, this means that coherent oscillations in the 
system's dynamics are about 1000 times faster than the spontaneous emission. We  simulate  several scenarios of coherent oscillation dynamics and show some  striking  results regarding  the physical implementation  of local as well as nonlocal gates useful for small-scale quantum computing based on the dimers.

\subsection{Swap gate and natural entanglement}
\label{dimergates}

The dimer can `naturally' generate the swap gate, which flips the two intermediate states of the 4-dimensional basis: $\ket{01}\rightarrow\ket{10}$, and 
vice versa. The matrix representation of the swap gate reads
\begin{eqnarray}
\label{SWAP}
	U_{\mathrm{swap}}=
	\left(
	\begin{array}{cccc}
	1 & 0 & 0 & 0 \\
	0 & 0 & 1 & 0 \\
	0 & 1 & 0 & 0 \\
	0 & 0 & 0 & 1 \\
	\end{array}
	\right) .
\end{eqnarray}
\begin{figure}[h]
  \centering
  \includegraphics[width=\columnwidth]{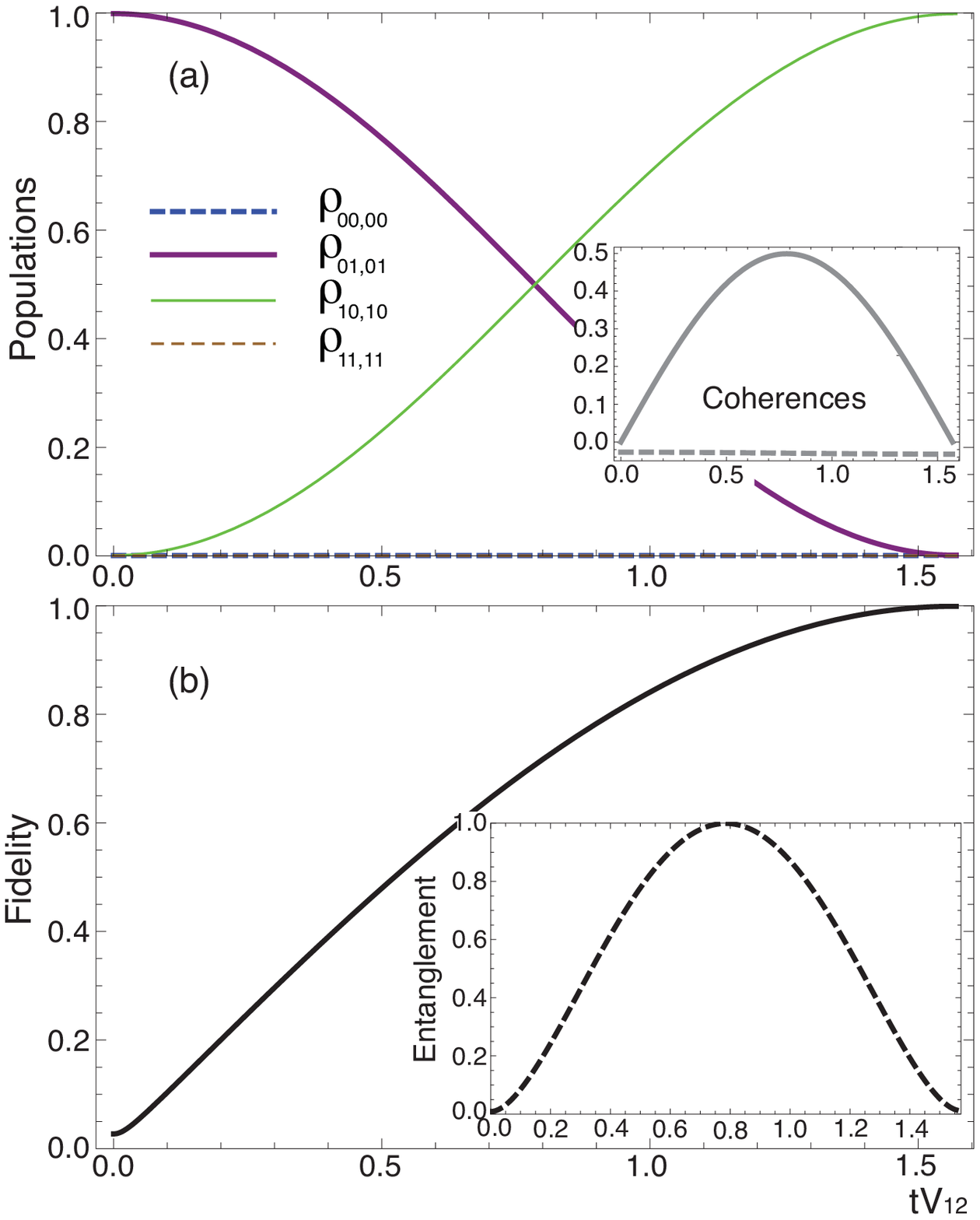}
  \caption{Natural swap gate dynamics. (a) Populations $\rho_{01,01}$ (solid-purple) and $\rho_{10,10}$ (thin-solid-green). 
   $\rho_{00,00}$ and $\rho_{11,11}$ are exactly zero. The inset shows $\mathcal{R}e\left[\rho_{01,10}\right]$ (dashed-gray) 
   and $\mathcal{I}m\left[\rho_{01,10}\right]$ (solid-gray) of the relevant coherence. (b) Main: Fidelity of the swap gate; the time of the gate is $t_{\mathrm{swap}}=\pi/2V_{12}$. Inset: evolution of the EoF 
   in the swap-gate process. $V_{12}=1356$~GHz, 
   $\Delta_-=14.3$~GHz, 
   $\Gamma=172$~MHz, and $\Gamma_{12}=-86$~MHz. The time is given in units of $V^{-1}_{12}$.
  }
  \label{fig1}
\end{figure}
Figure \ref{fig1} shows the pure generation of the swap gate for the situation $V_{12}/\Delta_- = 95 \gg 1$, see the caption for the detailed parameters. Note that the time axis has been plotted in $V^{-1}_{12}$ units. 
From the ground $\ket{00}$ state, if we computationally flip qubit 2 to its excited state, the dimer is driven to the 
$\ket{01}$ state. Then, under the action of the electronic coupling $V_{12}$, after a time $t_{\mathrm{swap}}=\pi/(2V_{12}) \sim 1.2$~ps
 the dimer reaches the $\ket{10}$ state, as shown in the main graph of Fig. \ref{fig1}(a) where we plot the populations of this evolution, as well as the dynamics of the coherence $\rho_{01,10}$ (inset). 
Figure \ref{fig1}(b) gives the corresponding fidelity $\mathcal{F}(\rho,\sigma)=\mathrm{Tr}\left[\sqrt{\sqrt{\sigma}\rho\sqrt{\sigma}}\right]$, 
where $\sigma$ is taken to be the expected state at the end of the gate and $\rho$ is the evolving state of the dissipative dynamics. We find that the swap gate step has been carried out within $\sim$~1.2~ps with $\mathcal{F} = 1$. We remark  that the swap gate operation continues (its dynamics exhibits coherent oscillations) for times up to two orders of magnitude longer than $t_{\mathrm{swap}}$. Intriguingly, the coherent oscillations lose only $5\%$ of the maximum fidelity after around $t=250 \times t_{\mathrm{swap}}$, i.e., for $t\approx 290$ ps (not shown).

One important byproduct of this conditional dynamics arises: 
By looking at the inset of Fig.~\ref{fig1}(b), it is clear that the swap gate can be tailored to generate entanglement in the dimer. The entanglement is quantified by the entanglement of formation (EoF). For two-qubit systems, the EoF is analytically computed as $EoF(\rho)=h\left(\frac{1+\sqrt{1-C^2(\rho)}}{2}\right)$, where $h(x)=-x\log_2x-(1-x)\log_2(1-x)$ 
is the binary entropy and $C(\rho)={\mathrm{max}}\{0, \lambda_1- \lambda_2- \lambda_3-\lambda_4\}$ is the so-called concurrence. 
$\lambda_i$'s are the eigenvalues of the matrix
$\sqrt{\rho(\sigma_{y}\otimes\sigma_{y}) \bar\rho (\sigma_{y}\otimes\sigma_{y})}$, where $\bar\rho$ is the 
elementwise complex conjugate of $\rho$~\cite{wootter}. The maximal value of entanglement is reached at the time $t_{\mathrm{swap}}/2$.  
This simple scenario clearly shows the versatility of the dimer to generate single- as well as two-qubit quantum gates.
\begin{figure}[h]
  \centering
  \includegraphics[width=\columnwidth]{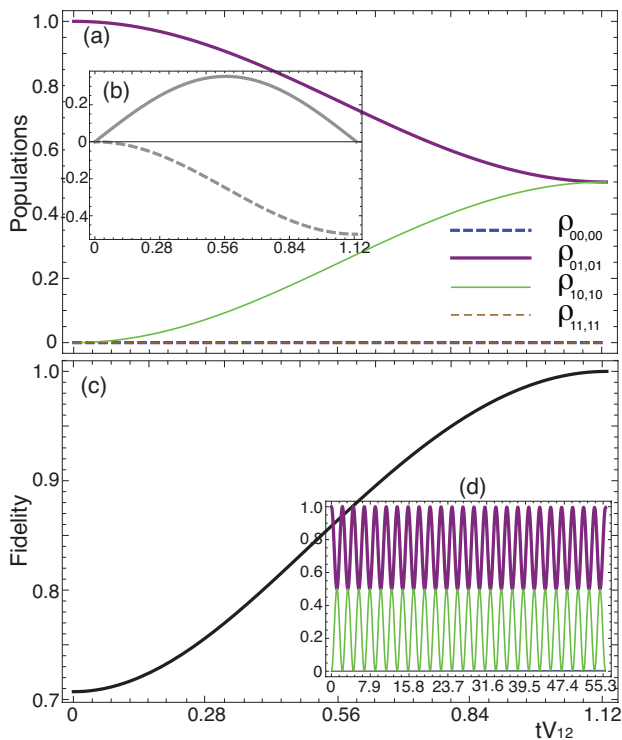}
  \caption{Natural generation of the maximally entangled Bell state $\ket{\Psi^-} = \frac{1}{\sqrt{2}}(\ket{01}-\ket{10})$, from the initial $\ket{01}$ state. (a) Populations, 
  (b) coherence $\mathcal{R}e\left[\rho_{01,10}\right]$ (dashed) and $\mathcal{I}m\left[\rho_{01,10}\right]$ (solid), (c) fidelity with respect to the ideal Bell state, and (d) populations for $50\times t_{\Psi^-}$. 
  $\Delta_-=190\times 14.3$~GHz 
  (see Fig.~\ref{fig1}). Remaining parameters are as in Fig.~\ref{fig1}.}
  \label{fig2a}
\end{figure}

We next focus on the naturally-generated entanglement by means of the swap gate. In Fig.~\ref{fig2a} we explicitly show the kind of entangled state that has been created for a ratio $V_{12}/\Delta_{-}=0.5$. It is possible to 
generate the antisymmetric Bell state $\ket{\Psi^-}=\frac{1}{\sqrt{2}}(\ket{01}-\ket{10})$ as shown by the populations (Fig.~\ref{fig2a}(a)) 
and coherences (Fig.~\ref{fig2a}(b)). The time required to obtain this entangled state, with a high fidelity (Fig.~\ref{fig2a}(c)), is $t_{\Psi^-}=\pi/2 \sqrt{\Delta_-^2/4 + V_{12}^2} \sim 819$~fs, i.e., it is determined by the interplay between the molecular detuning and the electronic coupling.  Such an entangled state is naturally robust to dissipation effects arising from the matrix host of the system and can be reached with fidelities around $95\%$ for times longer than $1000 \times t_{\Psi^-}$. Figure~\ref{fig2a}(d) shows the population oscillations for a time $50 \times t_{\Psi^-}$. 
It is worth noting that this entanglement dynamics is carried out with a molecular detuning two orders of 
magnitude higher than that used in Fig.~\ref{fig1}, showing the large range with respect to the ratio 
$V_{12}/\Delta_-$ for which the PBI dimers are able to implement a conditional quantum dynamics and entanglement generation.

We emphasise that the swap gate cannot be implemented experimentally following the procedure outlined above. Owing to the optical diffraction limit of $\lambda_i/2 \sim 250$ nm, we cannot address single qubits within a dimer using an external laser. Only the entangled symmetric and antisymmetric Bell states $\ket{\Psi^\pm}=\frac{1}{\sqrt{2}}(\ket{01}\pm\ket{10})$, two of the eigenstates of the dimer's Hamiltonian in Eq. (\ref{hrwa}), are optically accessible by a single-photon transition. Moreover, the doubly excited state $\ket{11}$ can be excited by a two-photon process \cite{Hettich}. Since these transitions into Bell states appear at different frequencies, the state to be excited can be selected by appropriately tuning the laser frequency. Alternatively, it can be easily shown from the molecular geometry (Fig. 1) that the symmetric and antisymmetric Bell states possess a mutually orthogonal transition dipole moment, i.e., selection is also possible using the laser polarisation. To generate an excitation localised on a single qubit of the dimer, and thus to realise the swap gate, a suitable coherent superposition of the Bell states $\ket{\Psi^\pm}$ is required, which can be achieved by an appropriate choice of the frequency bandwidth and/or polarisation of the laser. The subsequent dynamics within the system will then occur as outlined above. This {\it indirect} local action of the laser 
(or {\it computational} flipping) can be mathematically included in the model of Eq.~\eqref{ME} by assuming a local action of the laser 
Hamiltonian $H_L$.
Finally, we note that in the molecular spectroscopy community the Bell states $\ket{\Psi^\pm}$ are known as Frenkel (or molecular) exciton states.  

\subsection{Generation of the full entangled Bell basis}
\label{dimerbell}

We have  shown that the PBI dimer can naturally generate the Bell states $\ket{\Psi^{\pm}}$ by means of their strongly coherent electronic coupling. As these two states are part of a complete 4-state orthonormal basis, the so-called Bell basis, they can be transformed, by computationally performing  local operations, to the other two Bell states $\ket{\Phi^{\pm}}:=\frac{1}{\sqrt{2}}(\ket{00}\pm\ket{11})$. Appendix \ref{dimerlaser} shows an alternative scenario of dimer entanglement.
\begin{figure}[th]
  \centering
  \includegraphics[width=\columnwidth]{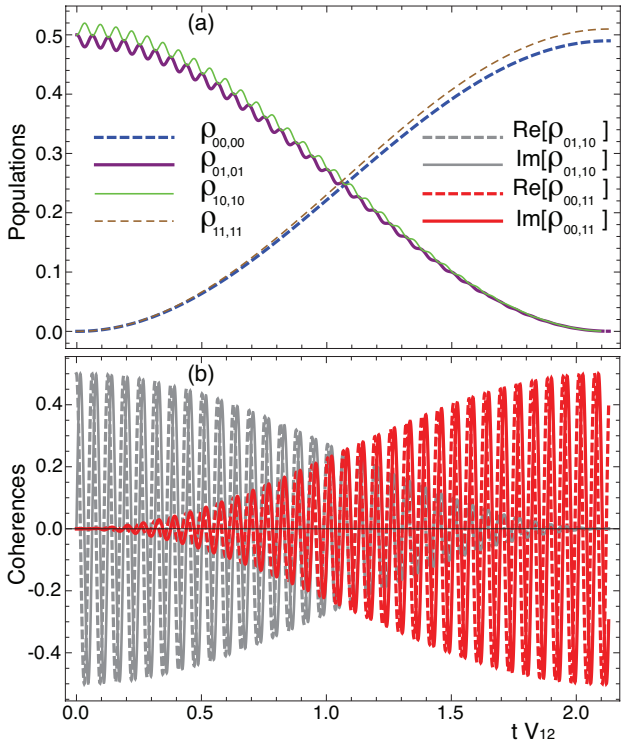}
  \caption{(a) Populations and (b) coherences $\rho_{01,10}$ (gray) $\rho_{00,11}$ (red) in the generation of $\alpha\ket{00}+\beta\ket{11}$ states from the 
  Bell $\ket{\Psi^{+}}$ state. Solid (dashed) stands for imaginary (real) part. The local action corresponds to flipping the state of qubit 2. Parameters as in Fig.~\ref{fig2a}.
  }
  \label{otherbellstate}
\end{figure}
\begin{figure*}[ht]
  \centering
  \includegraphics[width=14cm]{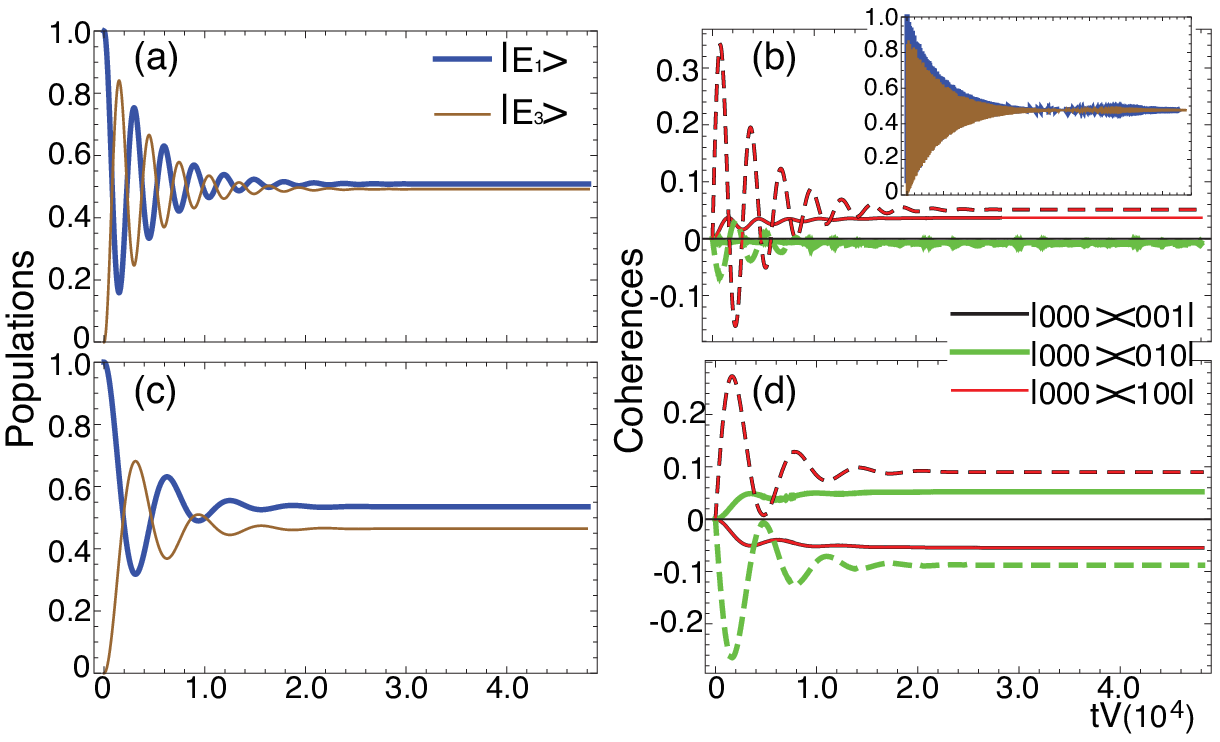}
  \caption{Expectation values for the transition 
  $\ket{E_1}\leftrightarrow\ket{E_3}$ under the action of a coherent (continuous) laser 
  $\Omega=1$~GHz. (a) Populations and (b) coherences (real (solid) and imaginary (dashed) part) with
  $\Delta_-=12000$~GHz and $\nu_L=E_3-E_1=700$~THz.
   (c) Populations and (d) coherences, $\Delta_-=1200$~GHz and 
  $\nu_L=E_3-E_1=699$~THz. 
  Real and imaginary curves for coherences $\rho_{000,001}$ (black) and $\rho_{000,100}$ (red) always take the same values, respectively. The inset shows the same 
  populations as in (a) but with a laser amplitude $\Omega=120$~GHz. Other used parameters are
  $V=1200$~GHz, $V_{13}=-120$~GHz, $\Gamma=172$~MHz, 
  $\gamma_{12}=\gamma_{23}=-86$~MHz, 
  and $\gamma_{13}=172$~MHz.
  }
  \label{figEE2}
\end{figure*}

As shown in Fig.~\ref{otherbellstate}, the initial state $\ket{\Psi^+}$ is driven to the state 
$\ket{\Phi'}:=\alpha\ket{00}+\beta\ket{11}$, with $\alpha\simeq0.70$ and $\beta\simeq0.57+0.42 i$. This can be 
done by computationally flipping qubit 2. The remaining matrix elements 
are at least two orders of magnitude smaller. A similar result is obtained if we start from the state $\ket{\Psi^-}$, 
in which case we arrive at  
$\ket{\Phi^*}:=-\beta^*\ket{00}+\alpha\ket{11}$ which in turn is orthogonal to the former one. 
This particular scenario has used the ideal 
$\ket{\Psi^{\pm}}$ states as our initial states: these can be prepared
by following the recipe in Fig.~\ref{fig2a} to entangle the monomers using the swap gate and then flipping the state of qubit 2. An alternative approach is illustrated in Fig.~\ref{fig2}.

The set of required quantum operations to generate, for example, the $\ket{\Phi^*}$ state (up to a global phase) can be concatenated as 
$\ket{\Phi^*}\simeq\mathbb{1}\otimes\sigma_x\; U_{\mathrm{swap}}\; \mathbb{1}\otimes\sigma_x\ket{00}$.
Although this three-gate circuit can be seen to be equivalent to the application of a local rotation on qubit 2 followed by a
controlled-NOT ($U_{\mathrm{CNOT}}^{12}$) operation,
we point out that the  dimer reported here is not able to directly simulate a controlled-NOT 
gate. This said,
we have shown that these PBI dimers allow us to naturally simulate the non-local swap gate, which, 
in conjunction with single qubit operations, implement  a universal gate set. 

We have already mentioned above that the Bell states $\ket{\Psi^{\pm}}$ can be experimentally generated by laser excitation. The other two Bell states  $\ket{\Phi^{\pm}}$ represent a superposition between the ground state and the two-photon accessible doubly excited state, which can also be induced by an external laser field.

\section{PBI trimer entanglement and nonlocality}
\label{trimercase}

We quantify the dynamics of the zig-zag-type trimer system (see Fig.~\ref{DDS}) by expanding the previous Hilbert space into 
the $2^{3}$ dimensional space  spanned by the computational basis states $\ket{i}\otimes\ket{j}\otimes\ket{k}$ ($i,j,k = 0,1$), 
taking into account all the cross-damping rates 
$\Gamma_{ij}$, and the coherent electronic couplings $V_{ij}$, 
which can  be directly computed from Eqs.~\eqref{CDR} and~\eqref{DCC}, by moving the 
subscripts $i,j=1,2,3$, and following a similar procedure to that for the dimer system. 

In Eq.~\eqref{trimersystem} we give the bare Hamiltonian (no laser) and in the Appendix~\ref{TQHam} the 
full (laser-driven) Hamiltonian for the PBI trimer, as well as the distance-dependence of the collective effects.
Diagonalisation of the Hamiltonian~\eqref{trimersystem} leads to the identification of three classes of eigenstates: (i) two product states, (ii) two purely pairwise 
entangled states, and (iii) four possible tripartite entangled states, see Eq.~\eqref{TEnergies}.  An example of the latter class is the state:
\begin{eqnarray}
	\ket{E_3}&=&\frac{2V}{\sqrt{2\Delta^-(\Delta^-+V_{13}-\Delta_-)}}\big(\ket{001}+\ket{100}\big)-\nonumber\\
	&& \sqrt{\frac{\Delta^-+V_{13}-\Delta_-}{2\Delta^-}}\, \ket{010} ,
	\label{eigs3}
\end{eqnarray}
with eigenenergy $E_3=-\frac{1}{2}(\nu-V_{13}+\Delta^-)$, 
where $\Delta^{\pm}=\sqrt{8V^2+(V_{13}\pm\Delta_-)^2}$. 
$\ket{E_3}$ is a pairwise entangled state if 
$V/\Delta_-\ll1$, but it exhibits genuine tripartite entanglement, otherwise.
The exact form of the eight eigenstates and their respective PBI trimer eigenergies 
are  left to the Appendix~\ref{BareHam}.

We can excite different transitions 
between the eigenstates by applying an external  coherent field. We begin by driving  the transition 
$\ket{E_1}\leftrightarrow\ket{E_3}$ with a weak laser ($\Omega=1$ GHz), 
and  assume as initial state $\ket{E_1}\equiv\ket{000}$. 
We first assume as specific case a ratio $V/\Delta_-=0.1$. Then the eigenstates are made up of pairwise entangled states and there are no tripartite entangled eigenstates. For instance, from Eq.~\eqref{eigs3} (see the numerics in Eq.~\eqref{TEnergies1}), it is clear that the intermediate eigenstate $\ket{E_3}$ has only $1.9\%$ of its population in the state $\ket{010}$, and almost all its population is in the superposition $0.70(\ket{001}+\ket{100})$. This means that the three PBI monomers are not entangled at the same time, but just two of them exhibit entanglement and their state is separable with respect to the other monomer. Under these conditions the transition $\ket{E_1}\leftrightarrow\ket{E_3}$ occurs coherently as shown by the time evolution of the expectation values $\bra{E_1}\rho(t)\ket{E_1}$ (blue) and $\bra{E_3}\rho(t)\ket{E_3}$ (brown) in Fig.~\ref{figEE2}(a) and (c). The stationary state is a statistical mixture of the two involved states. A similar result  is obtained when exciting the transition with a stronger laser amplitude $\Omega=120$ GHz (inset of Fig. \ref{figEE2}(b)). 

The situation differs when assuming a ratio $V/\Delta_-=1$ (Eq.~\eqref{TEnergies2}). In this case, the four 
intermediate eigenstates are reasonable superpositions of three orthonormal states and they exhibit 
tripartite entanglement (in fact, they all are W-like states). Such tripartite entanglement is present in the 
stationary regime being mixed with the ground state of the trimer (Fig.~\ref{figEE2}(c)) 
for the particular transition $\ket{E_1}\leftrightarrow\ket{E_3}$. 
The presence of some coherences at the end of the dynamics (more explicitly in Fig.~\ref{figEE2}(d)) implies that the stationary state is not completely classically correlated 
(a mixture of diagonal states) but still has quantum correlations assisted by the continuous action of the laser field.

\subsection{Natural entanglement dynamics}
\label{trimerentanglement}

\begin{figure}[h]
  \centering
  \includegraphics[width=\columnwidth]{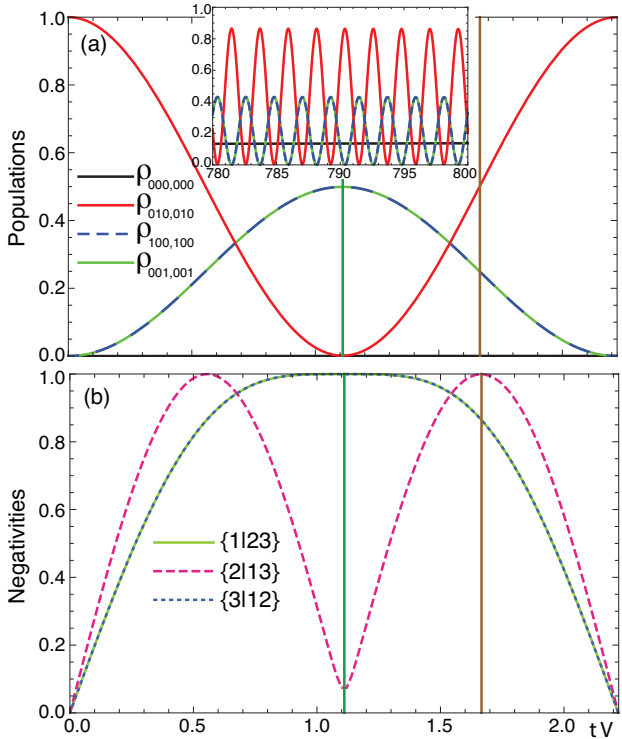}
  \caption{Generation of pairwise (bipartite) entangled and tripartite W-like states via the monomers dipole-dipole couplings. 
  (a) Main: Populations $\rho_{000,000}$ (black), $\rho_{010,010}$ (red), $\rho_{100,100}$ (blue) and $\rho_{001,001}$ (green). 
  Inset: Same populations for $tV\in[780\text{-}800]$. This is two orders of magnitude larger than the time in the 
  main plot and one order of magnitude shorter than the relaxation time. 
  Green vertical line at $t_{pw}$ indicates the generation 
  of a pairwise entangled state, and the brown line at $t_{\mathcal{W}}=3t_{pw}/2$ highlights the 
  generation of the $\ket{\mathcal{W}}$ state (see main text for full description). 
  (b) Negativity with respect to the partition $\{1|23\}$ (solid), $\{2|13\}$ (dashed) and $\{3|12\}$ (dotted). 
  Parameters: $V= 1356$~GHz, 
  $V_{13}= -122$~GHz ($|V_{13}|\sim0.09\, V$), 
  $\Gamma_{12}=\Gamma_{23}= -86$~MHz$,
  \Gamma_{13}= 172$~MHz, 
  and $\Delta_-=10$~GHz ($\sim0.007\, V$), 
  no laser. 
  $\nu_2>\nu_1=\nu_3$ (similar results are 
  obtained  for different choice of frequencies--not shown).}
  \label{Wstates}
\end{figure}

Pairwise as well as W-like tripartite entangled states~\cite{Acin2014,superactivation22013,jh17} are naturally generated as shown in Fig.~\ref{Wstates} if we initiate the trimer computationally 
in the $\ket{010}$ state, i.e., the sandwiched monomer (qubit 2) is in the excited state and the other two in their ground state. Note that in an experiment this state can only be excited by creating a suitable coherent superposition of eigenstates (see  Appendix~\ref{BareHam}).
After leaving the trimer to evolve exclusively by means of the electronic couplings, the system arrives to an almost perfect pairwise entangled state $1/\sqrt{2}(\ket{100}+\ket{001})\equiv\ket{\Psi^+}_{13}\otimes\ket{0}_2$ 
(see vertical green line in Fig.~\ref{Wstates}). 
This corresponds to a maximally entangled state between qubits 1 and 3. Hence, the trimer state is separable with respect to 
the second qubit.
This state is created after a time $t_{pw}=\pi/\sqrt{8V^2+(V_{13}-\Delta_{-})^2}\approx\pi/2\sqrt{2}\, V$, and such behaviour is expected according to the swapping effect due to $V$ and the experimental criteria  $V\gg V_{13}\geq\Delta_{-}$.  The $\rho_{100,100}$ and 
$\rho_{001,001}$ curves superpose each other, as seen in Fig.~\ref{Wstates}(a), and the inset shows the coherent dynamics of populations 
for a time frame two orders of magnitude larger than that in the main plot. 

Interestingly, this ultrafast dynamics allows the generation of tripartite entangled states as the 
so-called W-like states. Indeed, it is easy to show that under this evolution the only states propagating different from zero are those in the mixture $\rho_W(t)=p_1(t)\ket{000}\bra{000}+p_2(t)\ket{W^*(t)}\bra{W^*(t)}$, where 
$\ket{W^*(t)}=a_1(t)\ket{100}+a_2(t)\ket{010}+a_3(t)\ket{001}$, $|a_1(t)|^2+|a_2(t)|^2+|a_3(t)|^2=1$, and $a_2(0)=1$, 
are W-like states. Given the fact that $p_1(t)\ll p_2(t)$ for times shorter than 
the excited state lifetime, it follows that the states $\rho_{W}(t)\rightarrow\ket{W^*(t)}\bra{W^*(t)}$ are basically  the only ones present during the dynamics in this time frame. 

The entanglement of this evolution has been quantified via the negativity \cite{vidal2002} due to its operational interpretation 
and easiness of computation. For doing so, let us introduce $\{i|jk\}$ as a partition of the trimer system, where $i,j,k=1,2,3$ stand for 
qubits $1,2,3$ respectively. Hence, negativity is computed on the three partitions $\{1|23\}$, $\{2|13\}$, and $\{3|12\}$ 
as plotted in Fig. \ref{Wstates}(b). As expected, the negativities for $\{1|23\}$ and $\{3|12\}$ have the same behaviour due to the 
entanglement between qubits 1 and 3. 

It is clear that the representative $\ket{W}=\frac{1}{\sqrt{3}}(\ket{001}+\ket{010}+\ket{100})$ state belongs to the family of generated entangled $\rho_W(t)$ states. 
In Fig. \ref{Wstates}(a) such state occurs at the 
intersection of the three corresponding populations. Another scenario explores a non-trivial behaviour of the negativity for the 
partition $\{2|13\}$: At  its maximum value, marked by the brown line at 
$t_{\mathcal{W}}=(3/2) t_{pw}$ in Fig.~\ref{Wstates}, the trimer reaches the state 
\begin{equation}
	\ket{\mathcal{W}}=\frac{1}{2}\left(\ket{100}+\sqrt{2}e^{-\text{i} 0.489\pi}\ket{010}+\ket{001}\right).
	\label{wstsdc}
\end{equation}

This particular state is of great interest as it belongs to a subclass of W-like states that have been proven to be useful for teleportation 
and superdense coding \cite{agrawal2006}.

\subsection{PBI trimer nonlocal states}
\label{trimernonlocality}

So far we have discussed (Sections \ref{dimercoherence} and \ref{trimerentanglement}) the implementation of dimers and trimers based on PBI molecules as a valuable physical resource for quantum computing and information processing. We have also demonstrated conditional quantum dynamics and entanglement generation in dimers and trimers. 

In this section, our concern is whether the entangled states are also nonlocal states. Nonlocality~\cite{Acin2014,superactivation22013,jh17,andres17} is  a fundamental  feature of quantum states that is not always equivalent to entanglement~\cite{Acin2014} and has been demonstrated to 
be useful for some tasks in information theory~\cite{superactivation22013}. 
Nonlocality in bipartite states has been intensely studied and there are several different metrics for defining the nonlocality of quantum states, e.g.,  
CHSH inequality, activation, and super activation of nonlocality, just to name a few~\cite{superactivation22013,originalactivation2011,superactivation12012}. For 
more than two qubits, however, the nonlocality formalism extends to 46 classes of inequalities; 
each of them gives a classical limit that could 
be exceeded by nonlocal quantum states~\cite{ineq46}. Recently, analytical conditions to estimate the maximal violation of 
Mermin's inequality for three qubits were proposed~\cite{paul2016n,Adhikari2016}. Furthermore, an interesting development on 
multipartite nonlocality with operational (experimental) interpretation and implementation, in terms of 
inequalities that just involve one- and two-body expectation values (up to two parties correlations), has been reported~\cite{tura}.

Some of the eigenstates of the trimer's Hamiltonian 
support entanglement. They can be directly excited by a coherent laser and exhibit a robust 
dynamics against the slow dissipation due to spontaneous emission. The degree of entanglement, 
here caught by the Negativity (see Fig.~\ref{Wstates}(b)), depends on the interplay among the physical parameters; in particular, that between 
the molecular detuning and the effective electronic coupling. 

In the context of quantum nonlocality, a nonlocal state is an entangled one~\cite{chen}, 
but the opposite does not always hold, and there are plenty of  entangled but local states. As to the physical implementation of tripartite states, one question arises: are those trimer entangled 
states nonlocal? Here, we consider a Bell-like inequality and test it  for some trimer entangled states: if such an inequality is violated hence the corresponding state is said to be nonlocal.

 We numerically  test Mermin's inequality~\cite{mermin} by considering that two dichotomic observables act on each PBI monomer, hence  the inequality can be written as
\begin{eqnarray}
\label{BINQ}
 \Upsilon&\equiv&\big|\langle A_1B_2B_3\rangle+\langle B_1A_2B_3 \rangle + \\\nonumber
& &\langle B_1B_2A_3 \rangle- \langle A_1A_2A_3 \rangle\big|\leq2 ,
\end{eqnarray}
which is the so-called $(3,2,2)$ scenario: three parties, two observables per party, and two outcomes per observable (dichotomic observables),  and $\langle O \rangle=\mathrm{Tr}(\rho O)$ stands for the expectation value of the observable $O$. 
In our description of the PBI monomers as 
qubits, we write their associated  observables in terms of the Pauli matrices
$A_i=\cos{\theta_i}\sigma_z + \sin{\theta_i}\sigma_x$, and 
$B_i=\cos{\phi_i}\sigma_z + \sin{\phi_i}\sigma_x$, $i=1,2,3$. 
Other observables in terms of combinations with $\sigma_y$ can also be defined~\cite{Chang20091201}. 
However, as the states explored in this 
section have a matrix structure with their anti-diagonal elements identically zero, observables in terms of $\sigma_x$ plus $\sigma_y$ 
do not exhibit any violation. 
Hence, the inequality~\eqref{BINQ} is evaluated in terms of the different angles $\theta_i,\,\phi_i\in[-\pi,\pi]$, on 
the eigenstates supported by the trimer, and we search for at least one scenario in which Mermin's 
inequality~\eqref{BINQ} is violated. 

The eigenstate $\ket{E_3}$ (see Eq.~\eqref{eigs3}) transforms into the W-like state $1/\sqrt{3}(\ket{001}-\ket{010}+\ket{100})$ for the particular configuration 
$V=1200\,\mathrm{GHz},\,V_{13}=-120\,\mathrm{GHz},\,\mathrm{and}\,\Delta_-=1080\,\mathrm{GHz}$. According to Mermin's inequality~\eqref{BINQ} this state is 
of course nonlocal with a maximum violation numerically found to be $\Upsilon\sim3.05$.

We now look into the nonlocality of some of the states generated in the bare dynamics shown in Fig.~\ref{Wstates}. 
At $t=0$, the initial (product) state $\ket{010}$ is of course local. However, at a later time, 
the pairwise entangled state reached at 
$t_{pw}\approx\pi/2\sqrt{2}V$ (green vertical line in Fig.~\ref{Wstates}) exhibits a maximum value
$\Upsilon\sim 2.8$. In a similar way,  the W-like state Eq.~\eqref{wstsdc} reached at $t_{\mathcal{W}}\approx3\pi/4\sqrt{2}V$ (brown vertical line 
in Fig.~\ref{Wstates}) also violates Mermin's inequality as the function $\Upsilon$ attains a maximum of $\sim 2.2$. 
We then conclude that these two states  naturally generated by the trimer are both nonlocal states in the 
sense of the $(3,2,2)$ scenario. It is worth noting that the above two states are not pure at all because, in both cases, 
there exists a contribution due to the ground $\ket{000}$ state, as expected. Then, they both can be written as 
$(1-p)\ketbra{000}{000}+p\ketbra{\Psi_{pw}}{\Psi_{pw}}$ and $(1-p)\ketbra{000}{000}+p\ketbra{\mathcal{W}}{\mathcal{W}}$, 
respectively. We have identified $\ket{\Psi_{pw}}\equiv\ket{\Psi^+}_{13}\otimes\ket{0}_{2}$. Despite this fact, and thanks to the 
slow spontaneous emission of the trimer, the contribution of the ground state is up to three orders of 
magnitude smaller than the contribution of the relevant states. As a consequence, the maximum values obtained for the violation 
of the Mermin's inequality agree with the maximal violation of the corresponding pure state ($p=1$ in both cases).
This behaviour persists up to hundreds of picoseconds as it is shown for the bare dynamics in the 
inset of Fig.~\ref{Wstates}(a).

\section{Summary}
\label{summary}

For the implementation  of quantum logic gating, entanglement, and nonlocality in nanostructures based on organic molecules, we have considered here, without loss of generality, the particular arrangement shown in Fig.~\ref{DDS}. The transition dipole moments of the PBI molecules in the dimer and trimer span one plane and possess an opening angle $\theta=120^{\circ}$.
However, as mentioned above, the separation between the molecules (i.e. transition dipole moments) as well as their mutual orientation can be tailored by chemical synthesis. Hence, the values for the collective damping~\eqref{CDR}  and electronic couplings between transition dipole moments~\eqref{DCC} can be tuned in this way. For instance, the molecules could be arranged such that their transition dipole moments are parallel to each other; this would result in a smaller nearest-neighbour distance, thus in a stronger electronic coupling
and as a consequence in a higher degree of entanglement between them. 

For the dimer, we have shown how to drive a conditional quantum dynamics to achieve one-qubit (one-PBI-monomer) and two-qubit 
gates. We also demonstrated that all the entangled Bell basis states can be experimentally implemented in the dimer. 

In the trimer analysis we have additionally tested the nonlocality of the naturally generated entangled states.
We have numerically shown that a W-like state can be exactly obtained for specific combinations of the 
coherent electronic couplings and the molecular detuning. 
Furthermore, we also computed the corresponding 
locality violation for the dynamically 
generated pairwise ($\ket{\Psi_{pw}}$) and W-like ($\ket{\mathcal{W}}$) states
(see Fig.~\ref{Wstates}).

Our results on entanglement generation in both dimers and trimers reveal that the dynamics in these systems is highly coherent on the sub-picosecond and 
picosecond time scales; the relaxation time of their excited states lies in the nanosecond scale. 
This means that quantum gate operations with a high fidelity (coherent operations) are carried 
out in the sub-picoseconds scale (10$^4$-10$^5$ times faster than their life time).

Our study can also be extended to many-body systems because organic molecules can 
be synthesised to self-assemble into micrometre-long, fibrillar structures containing up to 10$^4$ molecules~\cite{richard2015}. The very dense packing of molecules in such systems results in strong electronic coupling between their transition dipole moments and thus should allow for the formation of entangled states on a macroscopic scale.

\section*{Acknowledgements} 

J.H.R. and C.E.S acknowledge support by the Colombian Science, Technology 
and Innovation Fund-General Royalties System (Fondo CTeI-Sistema General de Regal\'ias) under contract BPIN 2013000100007, 
Universidad del Valle for partial funding (grant CI 7930), and Colciencias (grant CI 71003). 
We gratefully acknowledge Andr\'es Ducuara 
for fruitful discussions. C.E.S. thanks Colciencias for a Fellowship and Universidad de C\'ordoba (grant CA-097). R.H. acknowledges support from the Elite Network of Bavaria (ENB, \textit{Macromolecular Science}) and from the German Research Foundation (DFG) through project HI1508/3.

\appendix

\section{General expressions for the PBI  collective damping and the dipole-dipole coupling}
\label{colleffects}

The coherent coupling $V_{ij} $, and the cross-damping rate $\Gamma_{ij}$ for a sample of $N$ qubits are computed, respectively, as 
\cite{Ficekbook}:
{\small \begin{eqnarray}
\nonumber
	\label{CDR}
	\Gamma_{ij} &=&
	\frac{3}{2}\sqrt{\Gamma_i\Gamma_j}
	\Big\{ 
	\left[\hat{\bm{\mu}}_{i}\cdot  \hat{\bm{\mu}}_{j}
	-(\hat{\bm{\mu}}_{i}\cdot \hat{\mathbf{r}}_{ij})
	(\hat{\bm{\mu}}_{j}\cdot  \hat{\mathbf{r}}_{ij})\right] \frac{\sin z_{ij}}{z_{ij}}+
	\\\nonumber
	&& 
	[\hat{\bm{\mu}}_{i}\cdot  \hat{\bm{\mu}}_{j}
	-3 (\hat{\bm{\mu}}_{i}\cdot \hat{\mathbf{r}}_{ij})
	(\hat{\bm{\mu}}_{j}\cdot \hat{\mathbf{r}}_{ij})] 
	\Big( \frac{\cos z_{ij}}{z_{ij}^{2}}-\frac{\sin z_{ij}}{z_{ij}^{3}} \Big) \Big\}, 
\\
	& & \\
	\nonumber 
	\label{DCC}
	V_{ij} & = &
	\frac{3}{4}\sqrt{\Gamma_i\Gamma_j}
	\Big\{   
	\left[( \hat{\bm{\mu}}_{i}\cdot \hat{\mathbf{r}}_{ij} )
	(\hat{\bm{\mu}}_{j}\cdot \hat{\mathbf{r}}_{ij}) 
	-\hat{\bm{\mu}}_{i}\cdot  \hat{\bm{\mu}}_{j}\right] 
	\frac{\cos z_{ij}}{z_{ij}} + \\  \nonumber
	&& [ \hat{\bm{\mu}}_{i} \cdot  \hat{\bm{\mu}}_{j}
	-3(\hat{\bm{\mu}}_{i}\cdot \hat{\mathbf{r}}_{ij})
	(\hat{\bm{\mu}}_{j}\cdot \hat{\mathbf{r}}_{ij})]
	\Big( \frac{\cos z_{ij}}{z_{ij}^{3}}+\frac{\sin z_{ij}}{z_{ij}^{2}} \Big)
	\Big\},\\
\end{eqnarray}}
where $z_{ij}= nk_{ij}r_{ij}$, $n$ denotes the matrix refractive index,
$k_{ij}= \omega_{ij}/c$, and $\omega_{ij}= \pi(\nu_{i}+\nu_{j})$. $\bm{\mu}_{i}$ is the dipole transition 
moment and $\bm{r}_{ij}$ is the separation vector between the centres of the two monomers $i$ and $j$; 
$i,j=1,...,N$. Under the rotating wave approximation-RWA, we can simplify the notation to $\omega_{ij}\rightarrow2\pi\nu_0$ as the inequality 
$|\nu_i-\nu_j|\ll\nu_0$ holds for all pairs of subscripts $ij$.
\begin{figure}[]
  \centering
  \includegraphics[width=\columnwidth]{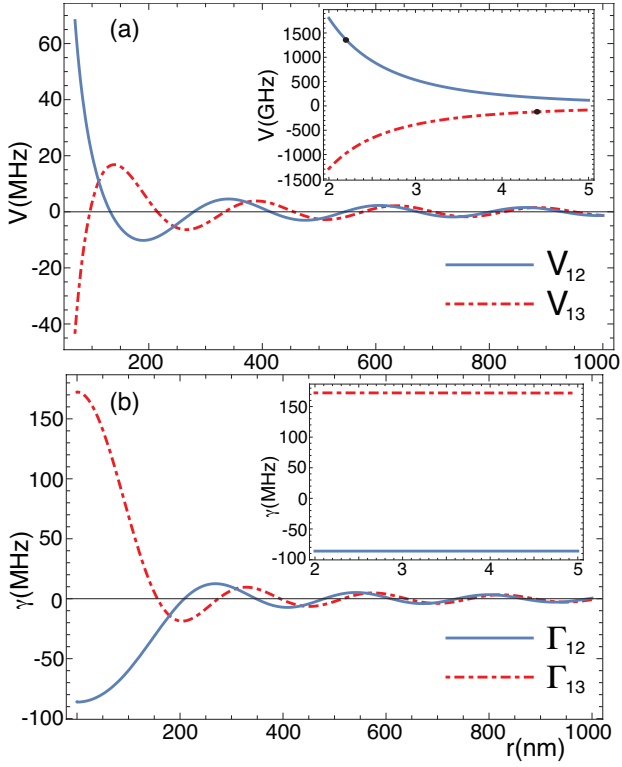}
  \caption{(a) 
  Dipole-dipole couplings $V_{12}$ (solid) and $V_{13}$ (dashed), (b) collective damping 
  $\Gamma_{12}$ (solid) and $\Gamma_{13}$ (dashed). The two black dots in the inset of (a) show the specific 
  inter-qubit separation for the computed values in the main text.}
  \label{CVDB}
\end{figure}

\section{PBI Trimer Hamiltonian}
\label{TQHam}

The driven Hamiltonian for the three-qubit system is straightforwardly extended from Eq.~\eqref{hrwa}. 
Considering the fixed 
coplanar configuration shown in Fig.~\ref{DDS} for the trimer, the Hamiltonian reads: 
\begin{eqnarray}
	\label{Threehrwa}
	&& \tilde{H}_{\mathrm{trimer}}= \\\nonumber
	&&\frac{1}{2} 
	\left(
	\begin{array}{cccccccc}
		-\delta_1  & \Omega & \Omega & 0 & \Omega & 0 & 0 & 0 \\
		\Omega & -\delta_2  & 2V_{23} & \Omega & 2V_{13} & \Omega & 0 & 0 \\
		\Omega & 2V_{23} & -\delta_3 & \Omega & 2V_{12} & 0 & \Omega & 0\\
		0 & \Omega & \Omega & \delta_2 & 0 & 2V_{12} & 2V_{13} & \Omega \\
		\Omega & 2V_{13} &2V_{12} & 0 & -\delta_2  & \Omega & \Omega & 0 \\
		0 & \Omega & 0 & 2V_{12} & \Omega & \delta_3 & 2V_{23} & \Omega \\
		0 & 0 & \Omega & 2V_{13} & \Omega & 2V_{23} & \delta_2 & \Omega \\
		0 & 0 & 0 & \Omega & 0 & \Omega & \Omega & \delta_1
	\end{array}\right) ,
\end{eqnarray}
where $\delta_1=3(\nu_0-\nu_L),\, \delta_2=\nu_2-\nu_L,\, \delta_3=\nu-\Delta_- -\nu_L$. Given the planar structure of the trimer (see Fig.~\ref{DDS}), 
we estimate the following values for the collective parameters; $V_{12}=V_{23}\approx 1356$~GHz,
$\Gamma_{12}=\Gamma_{23}\approx -86$~MHz
(the separation between monomers 1-2 and 2-3 is $2.2$~nm). 
For monomers 1 and 3 we have $V_{13}\approx -122$~GHz 
and 
$\Gamma_{13}\approx 172$~MHz 
as their separation is $4.4$~nm. For the specific computation of these values from the 
general expressions Eqs.~\eqref{CDR} and \eqref{DCC}, we have considered the dipole moments associated to 
qubits 1 and 3 to be parallel to each other, thus  the closer dimers exhibit a repulsive  interaction and the farthest 
ones an  attractive one.

Figure \ref{CVDB} shows the general behaviour of the collective parameters in the trimer system: (a) shows the behaviour of $V_{12}$ (solid curve) and $V_{13}$ (dashed curve) as functions of the mutual separation 
$\bm{r}$ from 
$70$~nm to $1000$~nm. The inset shows a zoom of such separation in the region $[2\text{-}5]$~nm. The inter-monomer separation for which 
the above numerical values were computed are 
represented by the two black dots in the inset of (a). The corresponding behaviour of $\Gamma_{12}$ 
and $\Gamma_{13}$ is shown in (b).

The time evolution of the trimer density matrix elements  are numerically computed by extending the master equation Eq.~\eqref{ME} to the new Hamiltonian Eq.~\eqref{Threehrwa}, and  by adding  $\Gamma_{3}$ and $\Gamma_{13}$ terms to the Lindblad operator.

\section{Laser-induced entanglement through doubly excited state}
\label{dimerlaser}
\begin{figure}
  \centering
  \includegraphics[width=\columnwidth]{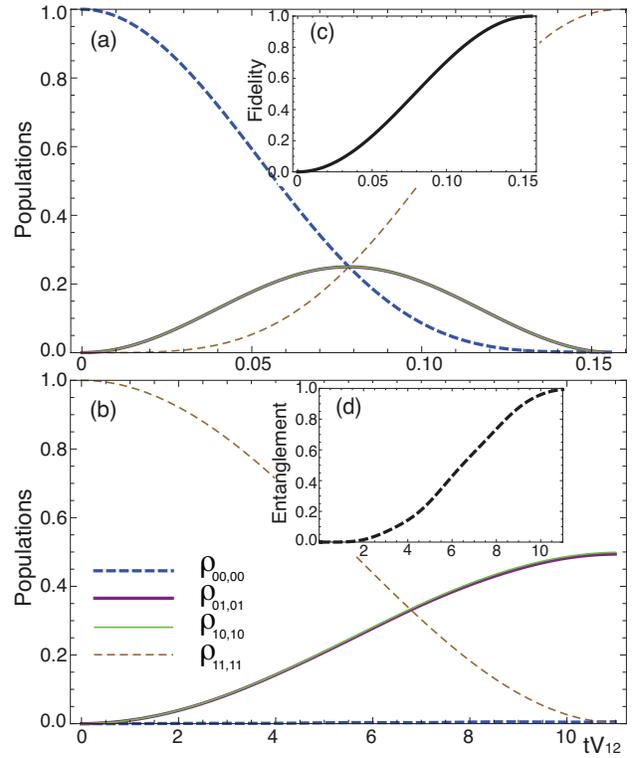}
  \caption{Entanglement from the doubly-excited state. (a) Transition from the ground $\ket{00}$ (dashed-blue) 
  to the doubly-excited $\ket{11}$ (thin-dashed-brown)
  state. $\Delta_+=0$ and $\Omega=27116$~GHz. 
  (b) Driven dynamics from the $\ket{11}$ state to the maximally entangled $\ket{\Psi^+}$ state. 
  $\Omega/2=135.6$~GHz, 
  and $\Delta_{+}=2\sqrt{(\Delta_-/2)^2+V_{12}^2}$. 
  (c) Fidelity evolution of the $\ket{00}\rightarrow\ket{11}$ transition. 
  (d) EoF generated during this process.
  $\Gamma$, $\Gamma_{12}$ and $V_{12}$ as in Fig.~\ref{fig1}, and the molecular detuning $\Delta_-=0.01 V$ (13.6~GHz).
  }
  \label{fig2}
\end{figure}

As an alternative of the natural entanglement generation shown in Section \ref{dimergates}, in this appendix we give another 
scenario in which the entangled $\ket{\Psi^{\pm}}$ states can be excited by means of a two-photon process. The dimer is 
driven to the doubly-excited $\ket{11}$ state within a time $\pi/\Omega$ ($\Omega\approx27116$~GHz), 
as shown in Fig.~\ref{fig2}(a).
This strong laser strength ($\Omega/2=10V_{12}$) 
is required as the energy difference between 
these two states is $\nu_1+\nu_2$. This transition occurs with high fidelity in $\sim 116$~fs (Fig.~\ref{fig2}(c)). 
After this step, the Bell state $\ket{\Psi^+}$ 
can be excited by setting the coherent laser to $\Omega/2=0.1V_{12}$ ($\Omega\approx271$~GHz),  
and applying it 
for a time  $t_{\Psi^+}\simeq7 \pi/10\Omega$ or, equivalently, for $t_{\Psi^+}\simeq7 \pi/2V_{12}$. 
This time roughly corresponds to $t_{\Psi^+}\sim 8.1$~ps (Fig.~\ref{fig2}(b)), and  the total process time is $\sim 8.2$~ps. 
Figure \ref{fig2}(d) shows the EoF for the second step.

In spite of the fact that the identical-molecule scenario ($\Delta_-=0$) would be a desired one, we have tested the more 
realistic case of detuned molecules. 
In doing so, we considered  $\Delta_-=0.01 V_{12}$ ($13.6$~GHz)
in Fig.~\ref{fig2}; however, we point out that this 
entanglement generation also works for $\Delta_-=0.1 V_{12}$, and even for $\Delta_-= V_{12}$ (1356~GHz):
in this latter case the maximum value for the EoF is $\sim0.85$. 
The interplay between the molecular detuning and the electronic interaction is 
also evident as the laser detuning must satisfy $\Delta_{+}=2\sqrt{(\Delta_-/2)^2+V_{12}^2}$ for this population transition to occur. 
If instead, we excite with a perfect resonance ($\Delta_+=0$), the ground state will increase quickly and an entangled state like that 
shown in Fig.~\ref{fig2}(b) will never appear. 
Additional to this resonance condition, a trade-off between the laser and the electronic interaction strengths is also a crucial factor for producing  the entanglement evolution of Fig.~\ref{fig2}(b) as they must satisfy $\Omega<V_{12}$. For laser strengths of the same order of or 
higher than $V_{12}$, the entangling effect is washed away. 

We emphasise that the PBI coherent dynamics persists  up to hundreds of picoseconds. For the case of 
Fig.~\ref{fig2}(b), the total mixed state is $(1-p(t))\proj{11}+p(t)\proj{\Psi^+}$, where the time evolution might be captured in the parameter 
$p(t)$, with $p(0)=0$, and hence for $t_m=m t_{\Psi^+}\equiv m (7 \pi/2V_{12})$; $m=1,2,3,...$, we have $p(t_{m})=1$, and 
thus the entangled $\ket{\Psi^+}$ state. 

\section{PBI Trimer eigensystem}
\label{BareHam}

The eight eigenstates of the bare trimer  Hamiltonian~\eqref{trimersystem} with their respective eigenenergies can be analytically computed and 
read
\begin{widetext}
\begin{small}
\begin{eqnarray}\label{TEnergies}
&&E_1=-\frac{3}{2}\nu_0;\; \ket{E_1}=\ket{000};\;\; E_2=-(\nu_2/2+V_{13});\; \ket{E_2}=\frac{1}{\sqrt{2}}(-\ket{001}+\ket{100});\nonumber \\
&&E_3=-\frac{1}{2}(\nu-V_{13}+\Delta^-);\; \ket{E_3}=\frac{2V}{\sqrt{2\Delta^-(\Delta^-+V_{13}-\Delta_-)}}(\ket{001}+\ket{100})-\sqrt{\frac{\Delta^-+V_{13}-\Delta_-}{2\Delta^-}}\ket{010};\nonumber\\
&&E_4=-\frac{1}{2}(\nu-V_{13}-\Delta^-);\; \ket{E_4}=\frac{2V}{\sqrt{2\Delta^-(\Delta^--V_{13}+\Delta_-)}}(\ket{001}+\ket{100})+\sqrt{\frac{\Delta^--V_{13}+\Delta_-}{2\Delta^-}}\ket{010};\nonumber \\
&&E_5=\frac{1}{2}(\nu+V_{13}-\Delta^+);\; \ket{E_5}=\frac{2V}{\sqrt{2\Delta^+(\Delta^++V_{13}+\Delta_-)}}(\ket{011}+\ket{110})-\sqrt{\frac{\Delta^++V_{13}+\Delta_-}{2\Delta^+}}\ket{101};\nonumber\\
&&E_6=\frac{1}{2}(\nu+V_{13}+\Delta^+);\; \ket{E_6}=\frac{2V}{\sqrt{2\Delta^+(\Delta^+-V_{13}-\Delta_-)}}(\ket{011}+\ket{110})+\sqrt{\frac{\Delta^+-V_{13}-\Delta_-}{2\Delta^+}}\ket{101};\nonumber\\
&&E_7=\nu_2/2-V_{13};\; \ket{E_7}=\frac{1}{\sqrt{2}}(-\ket{011}+\ket{110});\;\; E_8=\frac{3}{2}\nu_0;\; \ket{E_8}=\ket{111} ,
\end{eqnarray}
\end{small}
\end{widetext}
where $\Delta^{\pm}=\sqrt{8V^2+(V_{13}\pm\Delta_-)^2}$. 
To estimate the magnitude of the eigenenergies and the eigenstates  coefficients, 
we choose the following parameters: 
$V/\Delta_-=0.1$; $V=1200$~GHz, $V_{13}=-120$~GHz, $\Delta_-=12000$~GHz, 
$\nu=700$ THz and $\nu_2=712$~THz ($\Delta_-/\nu_0\simeq0.02$). Thus, the 
eigensystem now reads
\begin{widetext}
\begin{eqnarray}\label{TEnergies1}
&&E_1=-1056\,\mathrm{THz};\; \ket{E_1}=\ket{000};\;\; E_2=-355.9\,\mathrm{THz};\; 
\ket{E_2}=\frac{1}{\sqrt{2}}(-\ket{001}+\ket{100});\nonumber\\
&&E_3=-356.3\,\mathrm{THz};\; 
\ket{E_3}=0.7005(\ket{001}+\ket{100})-0.1361\ket{010};\nonumber\\
&&E_4=-343.8\,\mathrm{THz};\; 
\ket{E_4}=0.0962(\ket{001}+\ket{100})+0.9907\ket{010};\nonumber \\
&&E_5=343.8\,\mathrm{THz};\; 
\ket{E_5}=0.0981(\ket{011}+\ket{110})-0.9903\ket{101};\nonumber\\
&&E_6=356.1\,\mathrm{THz};\; 
\ket{E_6}=0.7003(\ket{011}+\ket{110})+0.1387\ket{101};\nonumber\\
&&E_7=356.1\,\mathrm{THz};\; 
\ket{E_7}=\frac{1}{\sqrt{2}}(-\ket{011}+\ket{110});\;\; E_8=1056\,\mathrm{THz};\; \ket{E_8}=\ket{111}.
\end{eqnarray}
\end{widetext}

For a smaller molecular detuning such that $V/\Delta_-~=~1$; $\Delta_-=1200$ GHz and $\nu_2=701$ THz, 
the eigensystem becomes 
\begin{widetext}
\begin{eqnarray}\label{TEnergies2}
&&E_1=-1050.6\,\mathrm{THz};\; \ket{E_1}=\ket{000};\;\; E_2=-350.5\,\mathrm{THz};\; 
\ket{E_2}=\frac{1}{\sqrt{2}}(-\ket{001}+\ket{100});\nonumber\\
&&E_3=-351.9\,\mathrm{THz};\; 
\ket{E_3}=0.5836(\ket{001}+\ket{100})-0.5646\ket{010};\nonumber\\
&&E_4=-348.2\,\mathrm{THz};\; 
\ket{E_4}=0.3992(\ket{001}+\ket{100})+0.8254\ket{010};\nonumber \\
&&E_5=348.2\,\mathrm{THz};\; 
\ket{E_5}=0.4174(\ket{011}+\ket{110})-0.8072\ket{101};\nonumber\\
&&E_6=351.7\,\mathrm{THz};\; 
\ket{E_6}=0.5708(\ket{011}+\ket{110})+0.5902\ket{101};\nonumber\\
&&E_7=350.7\,\mathrm{THz};\; 
\ket{E_7}=\frac{1}{\sqrt{2}}(-\ket{011}+\ket{110});\;\; E_8=1050.6\,\mathrm{THz};\; \ket{E_8}=\ket{111}.
\end{eqnarray}
\end{widetext}

As explained in the main text, the scenario given by Eq.~\eqref{TEnergies1} clearly shows that non tripartite 
entangled states are generated. Hence, only pairwise and product states build up the eigensystem. On the 
other hand, in the second scenario we can see that the states from $\ket{E_3}$ to $\ket{E_6}$ are superpositions with 
significant contributions around the three compounding states, thus being genuine tripartite entangled states.


\end{document}